\documentclass[aps,
 prb,
 graphicx,
 superscriptaddress,
 amsmath,
 amssymb,
 twocolumn,longbibliography]{revtex4-1}

\usepackage{graphicx}
\usepackage{bm}
\usepackage[utf8]{inputenc}
\usepackage[english]{babel}
\usepackage[T1]{fontenc}
\usepackage{mathptmx}
\usepackage{siunitx}
\usepackage{braket}
\usepackage{amsmath,amssymb}
\usepackage{hyperref}
\hypersetup{
    colorlinks=true,
    breaklinks=true,
    urlcolor=blue,
    linkcolor=blue,
    citecolor=blue}

\begin{document}
\title{Electronic materials with nanoscale curved geometries} 

\author{Paola Gentile}
\affiliation{CNR-SPIN c/o Universit\`a di Salerno, I-84084 Fisciano (Salerno), Italy}
\affiliation{Dipartimento di Fisica, Universit\`a di Salerno, I-84084 Fisciano (Salerno), Italy} 
\author{Mario Cuoco}
\affiliation{CNR-SPIN c/o Universit\`a di Salerno, I-84084 Fisciano (Salerno), Italy}
\affiliation{Dipartimento di Fisica, Universit\`a di Salerno, I-84084 Fisciano (Salerno), Italy} 
\author{Oleksii M. Volkov}
\affiliation{Helmholtz-Zentrum Dresden-Rossendorf, Institute of Ion Beam Physics and Materials Research, Bautzner Landstrasse 400, 01328 Dresden, Germany}
\author{Zu-Jian Ying}
\affiliation{School of Physical Science and Technology, Lanzhou University, Lanzhou 730000, China}
\author{Ivan J. Vera-Marun}
\affiliation{Department of Physics and Astronomy, University of Manchester, Manchester M13 9PL, United Kingdom}
\affiliation{National Graphene Institute, University of Manchester, Manchester M13 9PL, United Kingdom}
\author{Denys Makarov}
\affiliation{Helmholtz-Zentrum Dresden-Rossendorf, Institute of Ion Beam Physics and Materials Research, Bautzner Landstrasse 400, 01328 Dresden, Germany}
\author{Carmine Ortix}
\affiliation{Dipartimento di Fisica, Universit\`a di Salerno, I-84084 Fisciano (Salerno), Italy}
\affiliation{Institute for Theoretical Physics, Center for Extreme Matter and Emergent Phenomena, Utrecht University, Princetonplein 5, 3584 CC Utrecht, Netherlands}

\begin{abstract}
{\bf Research into electronic nanomaterials has recently seen a growing focus into the synthesis of structures with unconventional curved geometries including bent wires in planar systems and three-dimensional architectures obtained 
by rolling up nanomembranes. 
The inclusion  of these geometries 
has led to the prediction and observation of a series of novel effects that either result from shape-driven modifications of the electronic motion 
or from an intrinsic change of 
electronic and magnetic properties due to peculiar confinement effects. 
Moreover, local strains often generated by curvature also trigger the appearance of new phenomena due to the essential role played by electromechanical coupling in solids. 
Here we review the recent developments in the discovery of  these shape-, confinement- and strain-induced curvature effects at the nanoscale, and
discuss their potential use in electronic and spintronic devices.}
\end{abstract} 

\maketitle
\section{Introduction}
The electronic, magnetic and optical properties of materials acquire distinctive features when confined to two-dimensional sheets, one-dimensional nanowires and zero-dimensional quantum dots. As one progresses from an extended bulk solid to these nanostructures, size and quantum confinement effects become pervasive and strongly alter the material 
responses. 
In recent years, progresses in various nanostructuring techniques have enabled 
to expand electronic nanomaterials into the third dimension via 
the synthesis of 
nanoarchitectures that are 
made of constructs 
of two and one-dimensional nanomaterials. 
These advances have been largely triggered by the quest for high-density electronic memory and logic devices with a substantial increase in performance and reliability~\cite{par08,alb05,lav13}. 
Three-dimensional nanoarchitectures also address technological challenges in, for instance, three-dimensional magnetic sensing~\cite{bec19} and CMOS compatible magneto-impedance sensorics~\cite{kar15}. 

However, the realization of this new family of devices, which often require elements with curved geometric shapes such as spiralling tubes or helices,  also pose theoretical and experimental questions. 
First, can the curved geometry lead to novel variations of the fundamental properties of the material structure? 
Second, can these unprecedented physical properties be used to design electronic devices with different and even superior functionalities than the existing ones? 

At the fundamental level, 
a curved geometric shape introduces a new length scale crucial for the properties of the material structure: the characteristic radius of curvature. Consider for instance a diffusive electronic transport channel that is geometrically deformed. When the radius of curvature becomes comparable to the electronic mean free path, the charge and spin transport characteristics determined by the many (spin-flip) scattering events can be influenced by the bent electron trajectories imposed by the curved shape of the channel. 
Beside these classical shape effects on the electronic motion, 
curvature can also drive, via confinement and electromechanical coupling to inhomogeneous strain fields, strong deviations of the material intrinsic electronic properties. This occurs whenever 
the curvature radius approaches the de Broglie wavelength of the electrons near the Fermi level. 
In a similar manner, 
the ground state and the elementary excitations of materials exhibiting long-range order, like magnets and superconductors, can be geometrically tuned when the magnetic length or the superconducting coherence length are comparable to the curvature radius. 

Nanoscale geometry-induced effects can therefore pervade the physical properties of essentially all materials and endow them with
characteristics that would be 
hardly possible 
to achieve in conventional ``flat" structures of the same material. 
The focus of this Review is to highlight 
the physical origin of these new shape, confinement, and strain-induced functionalities, and how to exploit them 
in electronics, as well as in spin and even superconducting electronics.

\begin{figure*}
\includegraphics[width=\textwidth]{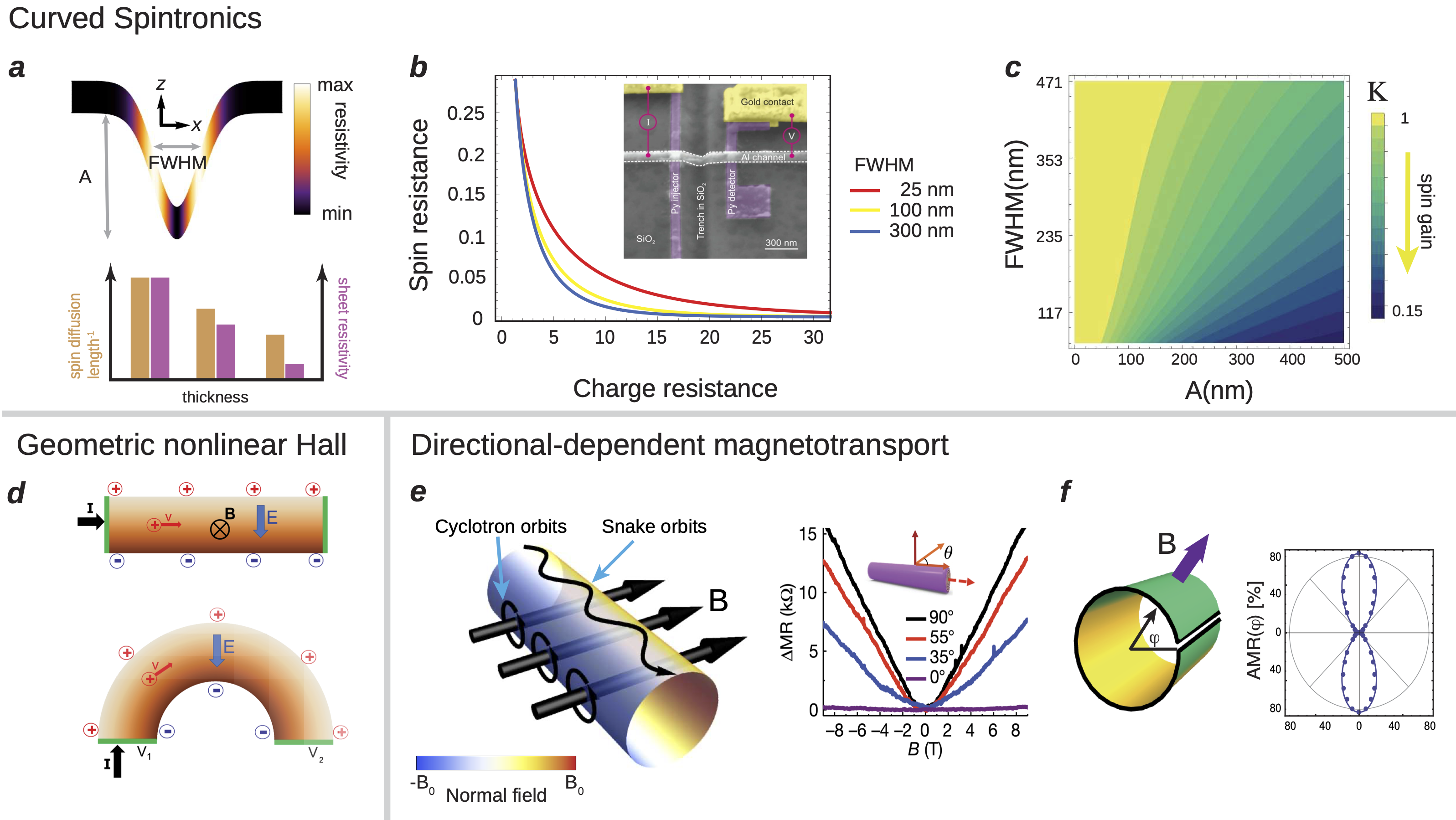}
\caption{{\bf Classical shape effects} (a) Sketch of an electronic channel grown on a trenched substrate with given height (A) and full width at half-maximum (FWHM). The thickness inhomogeneity along the channel creates a local resistivity distribution. There is a different scaling between the sheet resistivity $\rho(t)/t$ and the inverse of the spin diffustion length $\rho(t)$. 
(b) Room temperature spin resistance measured in a non-local spin valve ${\it vs.}$ the charge resistance of a curved Al channel grown on trenched substrates of different FWHM. The SEM image of the spin-valve is adapted from Ref.~\cite{das19}. 
(c) Color map of the curvature factor $K$ regulating the exponential decrease rate of the spin signal in a non-local spin valve with a trenched channel [see Ref.~\cite{das19}], and injector-detector distance $L=500$~nm.  Values of the curvature factor $K<1$ imply a gain in the spin signal relative to the charge resistance. 
(d) Sketch of the surface charges induced in the classical Hall effect, and the surface charges in a cuved wire that produce the electric field accelerating the carriers centripetally. 
(e) Color map of the normal component of a transversal magnetic field in a nanotube and the corresponding electronic semiclassical trajectories. The right panel shows the magnetoresistance measured in SnS$_2$/WSe$_2$ superlattices by tilting an axial magnetic field in the transversal direction [adapted from  Ref.~\cite{Zhao21a}]. 
(f) In carbon nanoscrolls with open geometries, a purely transversal magnetic field yields a strongly directional-dependent magnetoresistance [adapted from Ref.~\cite{cha17}].}
\label{fig1}
\end{figure*}  

\section{Classical Shape effects}
\subsection{Curved spintronics}
A first 
example of geometry-induced effects results from the structural inhomogenities that can be engineered in curved electronic channels.  Fabrication of 
curved channels
using a (non-planar) shaped substrate 
that acts 
as a template yields 
thicknesses 
with a local profile in one-to-one correspondence with the 
geometry. 
Specifically, and as evidenced in Fig.~\ref{fig1}(a), the thickness of a curved channel is strongly reduced in the regions of large 
curvature 
gradient. 
This characteristic can be used to efficiently tune the electrical properties of the system 
when the nanoscale thickness is comparable to the electronic mean free path. 
Even if quantum effects do not come into play due to the smallness of the de Broglie wavelength, classical size effects consisting in an  increase 
of the resistivity due to diffusive scattering at the channel and grain boundaries~\cite{ste02}, 
transform the shape-driven thickness inhomogeneities into an enhanced local nanoscale resistivity [see Fig.~\ref{fig1}(a)]. 

In addition, 
the local sheet resistance $\rho(t)/t$ 
decreases faster than the resistivity $\rho(t) \propto 1/t$ as the thickness $t$ increases [see Fig.~\ref{fig1}(a)]. 
This different scaling can be exploited in curved metallic nanochannels with engineered local thickness
of a few tens of nanometers when used as pure spin transport channels~\cite{jed01,val04,kim07}. 
Metallic materials form the basis of current spintronic technologies. In addition, 
the dominant spin relaxation mechanism corresponds to the so-called Elliot-Yafet mechanism~\cite{zut04,kim08}. This dictates that  the spin diffusion length is strictly locked to the resistivity of the metallic channel. The distinct scaling between local sheet resistance and resistivity consequently allows to control independently the charge and spin transport properties of a curved metallic channel. Such independent tuning has been realized~\cite{das19}
in lateral non-local spin valves [c.f. Fig.~\ref{fig1}(b)] . A given spin valve signal --  defined by the difference in the nonlocal resistance between parallel and antiparallel magnetic states of the injector and detector electrodes -- can be obtained for different values of the channel charge resistance [c.f. Fig.~\ref{fig1}(b)] and {\it vice versa}, in stark contrast with the case of conventional ``flat" channels where instead the spin and charge resistances are locked to each other. 
In addition, for devices with the same lateral footprint and at given charge resistance, the spin signal of a flat channel is always smaller than the electric response obtained using a curved channel. 
This is encoded in a curvature factor [c.f. Fig.~\ref{fig1}(c)] that quantifies the gain in the spin signal obtained 
using the intrinsic inhomogeneity of the curved channel.  
This generalized advantage combined with the independent tuning of charge and spin responses are 
of immediate relevance when considering practical implementation of spintronics: the use of geometric curvature to control on demand spin and charge impedances in multiterminal devices adds a novel approach for their efficient integration with complementary metal oxide semiconductor transistors. 

\subsection{Geometric nonlinear Hall effect} Geometric curvature can also lead to classical shape effects that directly derive from the charge motion in the tangential curved direction of the channel and can therefore appear even in planar structures. These are therefore completely different 
from the boundary scattering effect in the thickness direction at the basis of curved spintronics. 
In curved channels, charge carriers are forced to follow paths 
that in conventional flat channels typically require the presence of external electromagnetic fields.  
Consider for simplicity a planar curved wire taking the shape of a semicircular annulus~\cite{sch19}. Injection of a current in this channel necessarily leads to the appearance of a transverse electric potential: 
the current has to be accelerated radially to follow the circular path. 
This creates
surface charges similar in nature to those of the 
classical Hall effect  [c.f. Fig.~\ref{fig1}(d)] even if a perpendicular magnetic fields is absent. 
The ensuing transverse potential is quadratic in the current density, as required by time-reversal invariance, and can be thus regarded as a purely geometric nonlinear ``Hall" effect. The quadratic dependence on the injected current implies that an ac current with frequency $\omega$ will yield a transverse potential with $2\omega$ frequency. 
It is thus different than the transversal response occurring as a result of fluctuations in resistivity due to Joule heating since the latter gives rise to potential at higher harmonics. 
Using lock-in amplifiers, it is also possible to filter out other Hall-like contributions due, for instance, to spurious magnetic fields 
that yield potentials with $\omega$ frequency. Adopting this strategy, signatures of the geometric non-linear Hall effect have been individuated in graphene circular wires~\cite{sch19}. 

\subsection{Directional-dependent magnetotransport} The appearance of tortuous electronic trajectories in curved channels becomes even richer in the actual presence of external magnetic fields. This can be shown by considering the perhaps most common example of a nanostructure with curved geometry --  carbon nanotubes~\cite{book-CNT} -- in the presence of a transversal magnetic field.  
Charge carriers in carbon nanotubes respond primarily to the normal component of the externally applied magnetic field, which, as shown in Fig.~\ref{fig1}(e), changes sign at opposite side of the tube and averages out. 
When the cyclotron radius associated to the externally applied magnetic field $R_{cycl}= m^{\star} v_F / (e B)$,  with $v_F$ being the Fermi velocity and $m^{\star}$ a density-dependent dynamical mass~\cite{bha16}, is larger than the nanotube radius, the electronic trajectories correspond to helix-like orbits completely wrapping the tube. A direct computation of the magnetoconductance 
in the diffusive regime~\cite{cha17}
has revealed that these 
classical helical orbits yield, even in a single channel model, a quadratic longitudinal magnetoresistance experimentally observed in nanotube bundles~\cite{son94cnt} and multi-walled carbon nanotubes~\cite{cnt-exp}. 
In the regime where the cyclotron radius is smaller than the carbon nanotube radius, the nature of the classical electron trajectories changes qualitatively. The externally applied magnetic field is indeed large enough to allow for the formation of cyclotron orbits completely localized in the regions where the surface normal is parallel to the magnetic field~\cite{cre12}  [c.f. Fig.~\ref{fig1}(e)]. These cyclotron orbits do not contribute to the magnetoconductance contrary to the snake orbits that are instead naturally formed in nanotube regions where the 
normal component of the 
magnetic field changes its sign~\cite{2degsnake}. The snake orbits contribution to the magnetoresistance is characterized by 
a $\sqrt{B}$  power-law dependence~\cite{cha17}
explicitly shown in transport measurements~\cite{cnt-exp}. 
A similar  unconventional linear magnetoresistance has been also recently observed in a different material system: van der Waals SnS$_2$/WSe$_2$ heterostructures rolled-up into tubes~\cite{Zhao21a}. 
In this radial superlattice the magnetoresistance strongly decreases by tilting the magnetic field towards the axial direction [see the right panel of Fig.~\ref{fig1}(e)] thus suggesting the formation of snake orbits as responsible for this phenomenon. 

The imprint of snake orbits on the magnetotransport properties of tubular nanostructures becomes even more apparent when considering the open geometry of carbon nanoscrolls~\cite{nanoscroll, nanoscroll-nl} -- spirally wrapped graphite layers that unlike carbon nanotubes have overlapping fringes. In carbon nanoscrolls the density of snake orbits depends crucially  on the direction of the trasversal magnetic field. When the external magnetic field is directed towards the open edges there is a proliferation of snake orbits for the simple reason that the charge carriers feel an additional sign change of the effective normal magnetic field, as compared to the case in which the external magnetic field is directed orthogonal to the open edges. For instance in a single winding rolled-up open tube, the normal component of the magnetic field changes sign once or twice depending on the transversal magnetic field direction [c.f. Fig.~\ref{fig1}(f)]. 
Consequently, carbon nanoscrolls can display a strongly directional dependent magnetoresistance similar in its functional dependence to the anisotropic magnetoresistance of spin-orbit coupled magnetic materials. The magnitude of this effect, which is the immediate result of the broken rotation symmetry of the tubular structure, has been predicted~\cite{cha17} to be remarkably large in single winding carbon nanoscrolls as it can reach 80$\%$  [c.f. Fig.~\ref{fig1}(f)].  

\begin{figure*}
\includegraphics[width=\textwidth]{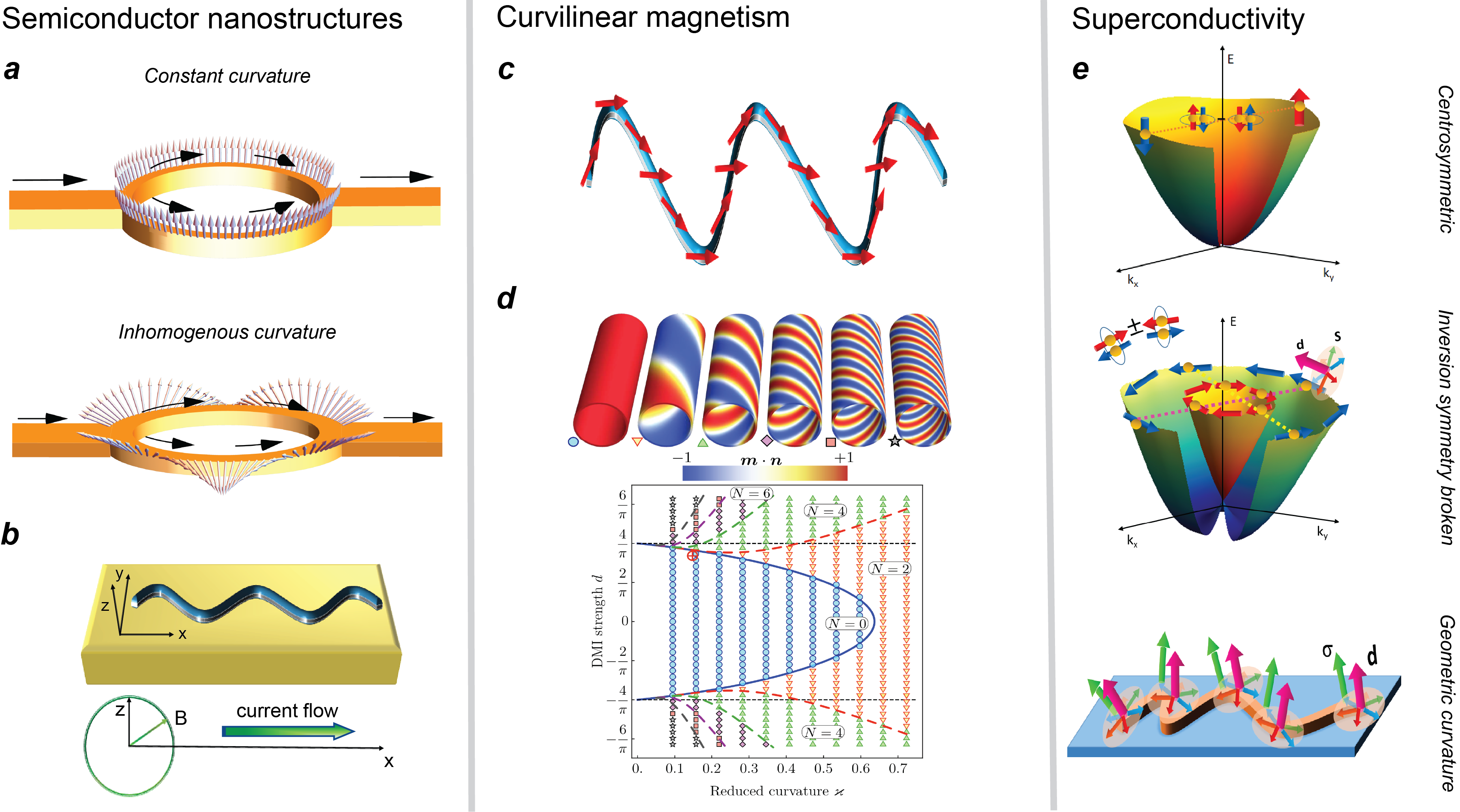}
\caption{{\bf Confinement-induced effects} (a) Schematics of the spin textures realized in circular (top) and shape-deformed (bottom) 
Rashba spin-orbit interferometers.
In the latter,  there is a spin-torque that leads to tangential spin components. 
(b) Top: Schematic view of  a semiconducting one-dimensional channel patterned in a serpentine shape at the mesoscopic scale. 
Bottom: Diagram of the pumping induced by the rotation of an external magnetic field. 
(c) Sketch of the magnetization textures (red arrows) along a bent magnetic wire with tangential anisotropy. 
(d) Equilibrium magnetic states in magnetic nanotubes with Dzyaloshinskii-Moriya interaction (DMI). Varying the strength of the DMI interaction and the curvature radius of the tube results in a phase diagram where the stable magnetic state is represented by the symbol associated to the nanotube [adapted from Ref.~\cite{yer20}].   
(e) Superconductors with center of inversion usually have an order parameter that is a spin-singlet state (top panel). In acentric materials, spin-singlet Cooper pairs are expected to be converted into a mixture of spin-singlet and spin-triplet states.These are characterized by a $\bm{d}$-vector that is generally parallel to the helical single electron spin $\sigma$. The spin $s$ of the Cooper pairs lies in the plane orthogonal to the $\bm{d}$ vector. In one-dimensional dimensional systems a texture of the $\bm{d}$ vector is generated in real space by curvature effects.}
\label{fig2}
\end{figure*} 
 
\section{Confinement-induced curvature effects}
 \subsection{Ballistic transport}
 Characteristic fingerprints of the formation of snake orbits can be also encountered in the ballistic transport regime of traditional semiconducting materials with mean free paths much larger than the curvature radii. An important example is represented by core-shell nanowires~\cite{core-nature}: systems consisting of a conducting, {\it e.g.} InAs, shell surrounding a, {\it e.g.} GaAs, core~\cite{rie12}. 
In their electronic band structure, the inhomogeneous radial component of the magnetic field leads to  Landau states condensed at small wavenumbers. 
 These states are connected for larger momenta to dispersive states corresponding in a semiclassical analysis to the snake orbits. 
 Such snake states are situated at the bottom of the energy spectrum. Consequently, they represent the main actors in the quantum transport at low chemical potential~\cite{snakenl}.
 Numerical calculations have shown that snake states result in peaks in the quantum conductance of core-shell nanowires, with the peak amplitude that can be tailored by adjusting the position of the metallic contacts with respect to the alignment of the transversal magnetic field. 
Similar conductance oscillations due to snake states have been also reported in graphene p-n junctions in the ballistic regime~\cite{pnsnake-nc1,pnsnake-nc2}. 

Semiconducting core-shell nanowires are however often grown with hexagonal cross sections and edges between different crystallographic facets that are rounded to form regions with finite curvature. In these regions, the corresponding local curvature radius can be small enough to be comparable to the de Broglie wavelength of the carriers, thus allowing to probe the quantum geometric potential arising from confinement on curved surfaces [see Box 1].  
This quantum geometric potential consists of a series of square wells~\cite{fer09} and thus makes the rounded edges of a prismatic core-shell nanowire regions of preferred localization. 
Such intrinsic curvature-induced localization has a strong interplay with the formation of snake orbits in the presence of a transversal magnetic field. Consider first 
a transversal magnetic field orthogonal to one of the facets of the core-shell nanowire: it will favour the formation of snake states on the two perpendicular edges enhancing their charge localization as compared to the other edges. 
On the other hand, for a magnetic field pointing toward one edge, snake states will form along the facets and counteract the tendency to localize the electronic charge at the nanowire edges. 
This different charge localization mechanism is then reflected in a ballistic directional-dependent magnetoresistance. 

An ideal three-dimensional topological insulator (TI) nanowire can be thought as of another material structure consisting of a conducting curved shell surrounding an insulating bulk. In the diffusive regime, the presence of an axial magnetic field gives rise to periodic Altshuler, Aronov and Spivak (AAS) magnetoconductance oscillations originating from weak antilocalization.  These AAS oscillations, with period $h/(2 e)$, are substituted in the (quasi)ballistic transport regime by $h/e$ periodic Aharonov-Bohm (AB) oscillations that are instead the result of the existence of the so-called perfectly transmitted mode. In curved geometries the single (spin-momentum locked) Dirac cone theory for the surface of a TI acquires a spin connection term that yields a Berry phase and consequently a gapped spectrum. However, a magnetic flux of half a flux quantum identically cancels the Berry phase and restores a gapless spectrum with an odd number of modes. Therefore, a perfectly transmitted mode with conductance $e^2 / h$ occurs~\cite{bar13}. AB magnetoconductance oscillations have been experimentally observed in Bi$_2$Se$_3$~\cite{duf13} and HgTe nanowires~\cite{zie18}. Superimposing a local variation of the nanowire curvature radius, {\it i.e.} considering nanocones, has been recently predicted~\cite{koz20} to lead to other intriguing mesoscopic transport phenomena such as resonant transmission through Dirac Landau levels.

\subsection{Spin-orbit coupled semiconducting materials}
Semiconducting nanomaterials 
also
exhibit
curvature effects that are due to relativistic corrections. The interplay between spin-orbit coupling and curvature leads to the appearance of complex spin-textures without counterparts in conventional ``flat" nanostructures. This is because in the rest frame of the electrons, the effective spin-orbit field has  a geometric component~\cite{nag13,yin16} [see Box 1] governed by the local geometric curvature. 
This geometric spin-orbit field 
has been directly probed in a simple class of nanostructures with curved geometries: semiconducting quantum rings with Rashba spin-orbit coupling~\cite{nit99,kon06}. 
In these systems, the out-of-plane tilt of the spin textures [see Fig.~\ref{fig2}(a)]
gives characteristic fingerprints in the conductance interference patterns~\cite{fru04,nag12}. 

The conductance modulations are regulated by the Aharonov-Anandan geometric phase -- the non-adiabatic analog of the Berry phase  -- which is in one-to-one correspondence with the spin textures~\cite{yin16}. Importantly, the quasi-periodic modulations can be additionally tuned by means of external planar magnetic fields.
The geometric 
spin-orbit field has remarkable consequences in geometries 
with inhomogeneous curvature~\cite{yin16}. 
In this case, the spin-orbit field exerts a torque on the electronic spin which then acquires a finite component parallel to the electron propagation direction [see Fig.~\ref{fig2}(a)]. 
The ensuing complex three-dimensional spin textures 
prove local curvature control of the Aharonov-Andan geometric phase and
can be directly probed in interferometric spintronic devices with unconventional geometries such as ellipses and squares~\cite{wan19}. 

The geometric spin torque 
has an important role 
also in 
zigzag-shaped nanowires where the local geometric curvature has a periodic profile~\cite{gen15}. The miniband structure of this 
geometric
superlattice is indeed characterized by the presence of minigaps opened by spin-orbit coupling at unpinned points in the mini Brillouin zone. The periodic buckling of a nanowire therefore induces a metal-insulator transition defining a geometric transistor switch. 
Additionally, the insulating states generally display Tamm-Shockley in-gap end modes~\cite{zak85}. 
The concomitant presence of the confinement-induced quantum geometric potential and the geometric spin torque can be also exploited to design new solid state electronic setups. For instance, it has been theoretically proposed that a zigzag-shaped nanowire with strong Rashba spin-orbit coupling can operate as a topological charge pump [see Fig.~\ref{fig2}(b)] in the complete absence of superimposed oscillating local voltages~\cite{pand18}, differently from the conventional pumping protocol in one-dimensional systems originally introduced by Thouless~\cite{tho83}. 
To operate the device, one uses 
an external rotating planar magnetic field -- generated for instance by running current pulses in two perpendicular conductors with a $\pi/2$ phase shift -- that serves as the periodic ac perturbation driving the charge pumping. The time-dependent Zeeman coupling acting on the spin textures realized by the geometric spin-orbit torque results ultimately in a sliding superlattice charge potential. Combining the latter with the charge gap opening mechanism provided by the quantum geometric potential leads to states
with a non-trivial Berry curvature in the synthetic two-dimensional mini-Brillouin zone, which, when integrated, yields a ``dynamical" non-zero Chern number~\cite{niu90}.
Hence, in each pumping period the device pumps two electronic charges with the quantization that is topologically protected against external perturbations, and can be relevant for metrological purposes. 

\subsection{Curvilinear magnetism} 
Curvature effects can be also exploited for novel spintronic device concepts relying on magnetic domain wall motion. In  magnetic wires with helical shapes, geometry offers unconventional means to control Rashba spin torque-driven~\cite{pyl16} as well as spin current-driven~\cite{yer16} domain wall dynamics. 
This is because 
the magnetic equilibrium state in thin films and wires is directly influenced by geometric curvature~\cite{gai14,she15}. For instance, 
in a buckled magnetic wire with tangential anisotropy
the magnetization vector generally displays local deviations from the 
tangential 
direction  [see Fig.~\ref{fig2}(c)].  
These effects are due primarily to the magnetic exchange 
energy,
which
 yields an effective Dzyaloshinskii-Moriya interaction (DMI) and an effective magnetic anisotropy [see Box 1]
 as a result of the confinement on curved geometries.  
 The existence of this effective antisymmetric exchange interaction
 has important consequences also on the existence and stability of modulated magnetic phases whenever the geometric curvature is comparable to the exchange magnetic length. Consider for instance a flat ultrathin film with an intrinsic DMI interaction. It is known that there exists a critical DMI strength separating homogeneous and periodic magnetization distributions~\cite{bog94}. If now the ultrathin film is bent with a constant curvature radius so as to create a magnetic nanotube, the effective curvature-induced DMI will renormalize the critical strength at which the modulated phase sets in~\cite{yer20}. In particular, by decreasing the curvature radius modulated phases can be stabilized by 
  the geometry 
 and appear even for vanishing intrinsic DMI strength [see Fig.~\ref{fig2}(d)]. Similar features are encountered 
  when considering skyrmions in magnetic nanotubes, which have been predicted to stably exist already in moderate magnetic field ranges~\cite{huo19}. 

Curvature effects yield peculiar features also in 
the dynamics of domain walls (DWs) in 
magnetic nanotubes. The configuration of a DW in a magnetic nanotube is characterized by the presence of a core-less vortex structure in the region separating the two oppositely magnetized domains.  
The two possible vorticities then distinguish two different DW configurations which are energetically degenerate but have different dynamical properties. Because of the presence of a finite radial magnetization component, an applied magnetic field exerts a torque on the DW. 
This, in turn, affects the radial magnetization itself. 
Whether the DW is distorted with a compression or enhancement of the radial magnetization depends on the handedness of the system defined by combining the vorticity of the DW with  the magnetic field vector direction~\cite{yan12dw}. 
This chiral dependent distortion 
ultimately modulates the motion of the DWs. 
Micromagnetic calculations based on the Landau-Lifshitz-Gilbert equation show that the velocity of the DWs is strongly dependent on the chirality. Importantly, also the stability of the DW displays chiral-dependent features. Certain chiral DWs have a completely suppressed Walker breakdown~\cite{sch74}: the collapse of the DW structure at a critical velocity, and one of the main complications for the optimization of memory device and logic gates based on fast and controlled DW motion. This is primarily due to the fact that breakdown of a DW in a tube involves a vortex-antivortex pair creation contrary to the single vortex-mediated breakdown of DW in flat thin films.
Based on this topological constraint, DWs generally have a strongly enhanced stability when guided in tubular nanostructures.  

Chiral symmetry breaking effects have been predicted~\cite{ota16} and more recently observed~\cite{kor21} in magnonics. Spin waves -- magnetic excitations that hold potential in 
information 
processing
--  acquire a peculiar asymmetric dispersion in magnetic nanotubes with an equilibrium state in which the magnetization rotates around the circumference of the tube. 
Specifically, spin waves of same frequency propagating in different directions are characterized by different wavelengths.  
Since the rotating magnetization together with the propagation direction defines a handedness, the occurrence of this phenomenon implies a chiral symmetry breaking. 
Importantly, this effect, occurring in conventional thin film geometries only in the presence of  intrinsic DMI interaction, 
does not originate from the exchange interaction bur rather from the non-local dipole-dipole interaction.

\subsection{Superconducting electronics}
Recent theoretical studies have suggested that 
the effective spin torques activated by geometry might affect also electronic pairing and possibly lead to new functionalities in
superconducting spintronics.
One of the paradigms of superconducting spintronics is to exploit spin-triplet Cooper pairs since they can carry angular momentum without energy dissipation, and thus can be functionalized to yield spin-polarized supercurrents for ultrahigh energy-efficient storage and transfer of information \cite{Linder15}. 
Spin-triplet pairing has a vectorial nature and can be typically encoded in the so-called ${\bm d}$-vector \cite{Sigrist91} whose components correspond to the zero spin projections of the triplet state along the symmetry axes [see Box 1]. 
A crucial challenge in the area of superconducting spintronics is to achieve control mechanisms and devise systems that are able to convert spin-singlet into spin-triplet electron pairs since spin-singlet superconductors are more abundant in nature. This issue has been largely investigated by engineering superconductor-magnet heterostructures with suitably designed non-collinear magnetic patterns~\cite{ber01,Eshrig08,Robinson10}
and more recently by considering spin-orbit coupling without breaking time-reversal symmetry \cite{ber14}. 
In this context, systems with Rashba spin-orbit coupling due to structural inversion asymmetry are particularly appealing.

In two dimensions the lack of inversion symmetry removes the spin degeneracy of the Bloch states and a spin texture develops in the Brillouin zone due to the Rashba interaction that couples the electron spin with the crystal wave vector \cite{gor01}.  
For Cooper pairs, 
this symmetry reduction forces the occurrence of a mixing of spin-singlet and spin-triplet pairing at the Fermi level. 
Moreover, due to the spin anisotropy, the spin-triplet ${\bm d}$ vector follows the electron spin orientation \cite{Frigeri04} resulting into a helical pattern [see Fig.~\ref{fig2}(e)]. 
If at a given momentum $k$ at the Fermi level the electron spin points along the $y$ direction, $\uparrow_y$, with the time reversal partner $\downarrow_y$ at $-k$, the resulting spin-triplet configuration can only be with vanishing total spin projection along the $y$ orientation thus corresponding to the $d_y$ component.  
In analogy to the spin-triplet texture occurring in momentum space along the Fermi line, a variation of the ${\bm d}$-vector orientation occurs in 
real space when geometric curvature is present \cite{Ying17}. 
As discussed above, the geometric spin-orbit torque yields local variations of the electron spin orientation \cite{yin16}. 
Hence, to avoid pair breaking, the spin-triplet configuration has to follow the spin anisotropy by twisting the ${\bm d}$ vector according to the curvature of the electronic channel [see Fig.~\ref{fig2}(e)]. 
Crucially, 
the reconstruction depends on the 
ratio between the 
geometric curvature and the superconducting coherence length. 
Then, the combined presence of geometric curvature and spin-orbit coupling has the effect to generate spin-triplet pairs with an orientation that is substantially dictated by the 
profile of the confining potential. These mechanisms can lead to striking effects in superconducting electronics. 
As a direct example, the Josephson effect can be mechanically controlled both in the amplitude and phase of the supercurrent by geometrically curving the superconducting electrodes \cite{Francica20}. 

\begin{figure*}
\includegraphics[width=\textwidth]{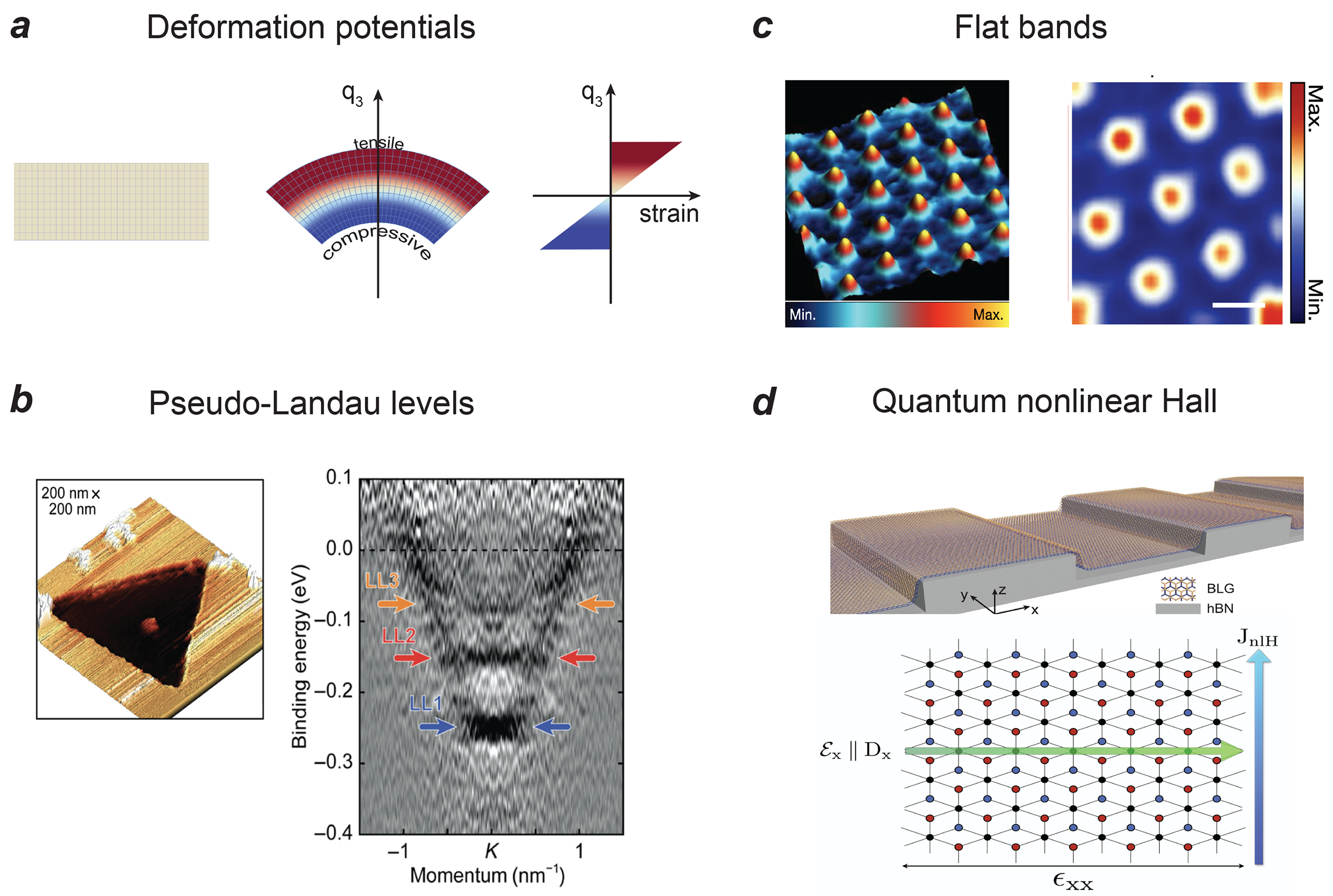}
\caption{{\bf Strain-induced effects} (a) Bending a flat nanostructure creates a strain gradient with regions under compressive and tensile strain separated by the mechanical neutral plane $q_{3} \equiv 0$.  
(b) Graphene grown on SiC substrates displays nanoprisms as evidenced in atomic force microscopy tomography where one Si layer is missing. The formation of pseudo-Landau levels in the strained nanoprisms can be observed even at room temperature using angle-resolved photoemission spectroscopy. Adapted from Ref.~\cite{nig19}.
(c) Left: STM topography. Right: Measured dI/dV maps at energy corresponding to one of the flat-band regions created by the buckling-induced periodic pseudo-magnetic field. Adapted from \cite{mao20}.
(d) Top: Artificially corrugated bilayer graphene (BLG) on a patterned hexagonal boron nitride substrate. Adapted from Ref.~\cite{ho21}.
Bottom panel: Top view of the regions where bilayer graphene is strained. A nonlinear Hall current J$_{nlH}$ is generated under the application of an electric field $E_x$ parallel to the dipole $D_x$. Adapted from Ref.~\cite{bat19}.
}
\label{fig3}
\end{figure*}

\section{Strain-driven curvature effects}

\subsection{Strain-induced geometric potential} 
In nanosystems with curved geometries electromechanical coupling leads to effects that often coexist and amplify the confinement-induced curvature effects discussed above. 
A key property of bent nanostructures fabricated using strain engineering methods 
is the presence of local strain fields varying on the nanoscale~\cite{mei07}. In insulators, the presence of these strain gradients is at the basis of the flexoelectric effect -- an electromechanical coupling that, contrary to piezoelectricity, is universal and symmetry-allowed in all solids~\cite{zub13}. 
Strain gradients spontaneously generated in curved nanostructures yield remarkable effects also in semiconducting materials. According to the linear potential deformation theory~\cite{wal89}, a local strain yields a shift of the conduction (valence) band edges, and therefore corresponds to a local potential attracting the charge carriers towards the maximally strained regions  [see Fig.~\ref{fig3}(a)]. This potential is therefore directly proportional to the local geometric curvature with a corresponding characteristic energy scale that lies in the eV range for conventional semiconductors. Using an adiabatic separation of fast and slow quantum degrees of freedom~\cite{ort11b}, it can be shown that this local potential ultimately yields a strain-induced geometric potential of the same functional form of the confinement-induced quantum geometric potential but strongly, often gigantically, boosting it. Such large enhancement of curvature effects renders the phenomena predicted in curved nanostructures, ranging from winding-generated bound states in rolled-up nanotubes~\cite{ort10} to topological band structures in systems constrained to periodic minimal surfaces~\cite{aok01}, observable above the sub-Kelvin energy scale. 
In micrometer sized wrinkled ribbons of GaAs, for instance, curvature-induced localized end states have been predicted to be observable up to a few Kelvin 
if local strain fields are explicitly taken into account~\cite{pan16}. 

\subsection{Gauge fields in Dirac materials}  
Strain fields yield remarkable effects also
 in Dirac materials, such as graphene. 
 One can distinguish between two different strain-induced effects. First, there is a 
renormalization of the Fermi velocity directly proportional to the local geometric curvature~\cite{dej07}. This space-dependent Fermi velocity is expected to induce 
spatial
oscillations of the local density states near the ripples 
that are naturally formed 
in suspended graphene samples~\cite{mey07}. 
The
correlation between the morphology of the graphene samples and their electronic properties 
could then explain the local variations of the charge compressibility 
observed in scanning single electron transistor experiments~\cite{mar08}.  

In Dirac materials, in-plane mechanical deformations~\cite{Juan2012} lead to another curvature-induced effect. 
Specifically, electrons can react to strain as if external electromagnetic fields were applied. 
Strain fields indeed result in effective gauge fields [see Box 2] that are opposite in the two graphene valleys. 
These gauge fields lead to a complete reorganization of the spectrum when they generate a ``pseudo"-magnetic field, {\it i.e.} a magnetic field that is opposite in the two valleys. The latter indeed leads to the appearance of pseudo-Landau levels, and thus represents an 
important example of strain engineering~\cite{gui10}. 
Pseudo-Landau levels have been directly imaged using scanning tunneling microscopy in graphene nanobubbles grown on a platinum surface~\cite{lev10}, as well in flakes supported on nano-pillars~\cite{jia17pseudo}.   
Nanobubbles, in particular, have been shown to generate pseudomagnetic fields as high as 300~T due to the large in-plain strains~\cite{lev10}. It is important to note that nanobubbles are generally present in van der Waals heterostructures but typically avoided in high-quality devices since in-plane strains represent a dominant  factor limiting the electronic mobility~\cite{cou14}.  However, nanobubbles have been recently used as active elements for the formation of nanometer-scale lateral p-n junctions in a charge-transfer graphene-based heterostructure~\cite{riz22}.

Strain-induced Landau levels have been also generated in triangular nanoprisms of a SiC  substrate [see Fig.~\ref{fig3}(b)] and observed  by 
angle-resolved photoemission spectroscopy~\cite{nig19}. 
A periodic arrangement of pseudomagnetic fields with periods in the tens of nanometer scale [see Fig.~\ref{fig3}(c)] has been instead realized in buckled graphene superlattices~\cite{mao20}.
The pseudo-Landau levels in this case form weakly dispersive bands that are strongly localized in real space [see Fig.~\ref{fig3}(c)]. 
Since the kinetic energy is quenched, the system can also develop a correlated phase characterized by a pseudogap-like depletion of the density of states, similar to the situation encountered in magic-angle twisted bilayer graphene~\cite{cao18}. 
Buckling instabilities can be thus exploited to investigate interaction phenomena with important advantages in ease of fabrication and scalability as compared to twistronics. 

\subsection{Quantum nonlinear Hall effect} 
Strain-induced gauge fields 
trigger curvature-induced phenomena 
not strictly related to the presence of pseudo-Landau levels. 
These occur in 
non-centrosymmetric materials with Berry curvature: the quantity that encodes the geometric properties associated to the quantum electronic wavefunctions. 
More specifically, these effects originate from the  interplay between the intrinsic crystalline anisotropies of the Dirac material and structural anisotropies induced by certain mechanical deformations. 
In their pristine form, two-dimensional Dirac materials have a trigonal crystalline structure. 
Suppose to either induce an highly anisotropic  one-dimensional buckling instability or to deposit the nanomembrane onto an anisotropic pre-patterned substrate [see Fig.~\ref{fig3}(d)]. 
The end product is a superstructured material with an unusually low crystalline symmetry, where all rotational symmetries are broken. 

Electronic systems with substantial Berry curvatures and such low-symmetry crystalline content can exhibit a segregation of positive and negative regions of Berry curvature in momentum space leading to a net dipole moment.
This Berry curvature dipole yields a quantum nonlinear Hall effect in time-reversal symmetry conditions~\cite{Sodemann2015,Xu2018,ma19}:  a non-linear electrodynamical phenomenon relevant for high-frequency rectification and long-wavelength photodetection~\cite{zha21}. 
Creating anisotropic mechanical deformations in gated bilayer graphene 
satisfies all the requirements for the existence of a sizable Berry curvature dipole. 
Inversion symmetry breaking is achieved with an external electric field perpendicular to the layers. 
Moreover, anisotropic strains~\cite{bat19} already in their simplest homogeneous form have been predicted to trigger Berry curvature dipoles [see Fig.~\ref{fig3}(d)]  comparable to the ones of transition metal dichalcogenides. 
The observation of a large Berry curvature dipole in corrugated bilayer graphene~\cite{ho21} has brought to reality 
the concept of such a curvature-induced quantum nonlinear Hall effect. Importantly, this also 
opens a new 
approach to
 electronic devices that can be used as energy harvesters and terahertz detectors via geometric design.

\section{Synthesis and characterization methods} 
The creation and exploration of curvature effects strongly relies on the ability to synthesise nanoscale objects with a targeted geometric shape. 
Curvilinear one-dimensional nano-objects in planar structures can be fabricated using state-of-the-art thin-film technology processing, including electron beam litography and ion-beam etching. 
These approaches have been widely used for the fabrication of, for instance, (quantum) rings~\cite{Klaui05a,nag12,nag13}, curved parabolic stripes~\cite{Volkov19c}, or square loops~\cite{wan19} of different materials. 
The main advantages of this approach is that the functional curvilinear nanostructure can be easily supplemented with stable electrical contacts for change injection and gate electrodes. 
Furthermore, the equilibrium spin textures of magnetic materials 
can be retrieved using a number of techniques ranging from Lorentz transmission electron microscopy~\cite{Phatak12} to X-ray magnetic circular dichroism photoelectron emission (XMCD-PEEM)~\cite{Streubel16a}. Imaging the magnetization states of Ni$_{81}$Fe$_{19}$ (permalloy) parabolic stripes using XMCD-PEEM has provided the first experimental evidence of the exchange-driven DMI coupling in curved geometries~\cite{Volkov19c}. 
Another advantage of planar structures is that  a unique sample can consist of arrays of nano-objects. This configuration is ideal in electronic transport since it naturally provides ensemble averaging that filters out device to device variations~\cite{nag12,nag13}. 
The use of InGaAs square loops arrays~\cite{wan19} for instance has been essential to identify distinctive signatures of the geometric spin-orbit torque in quantum conductance experiments. 

\begin{figure*}
\includegraphics[width=15cm]{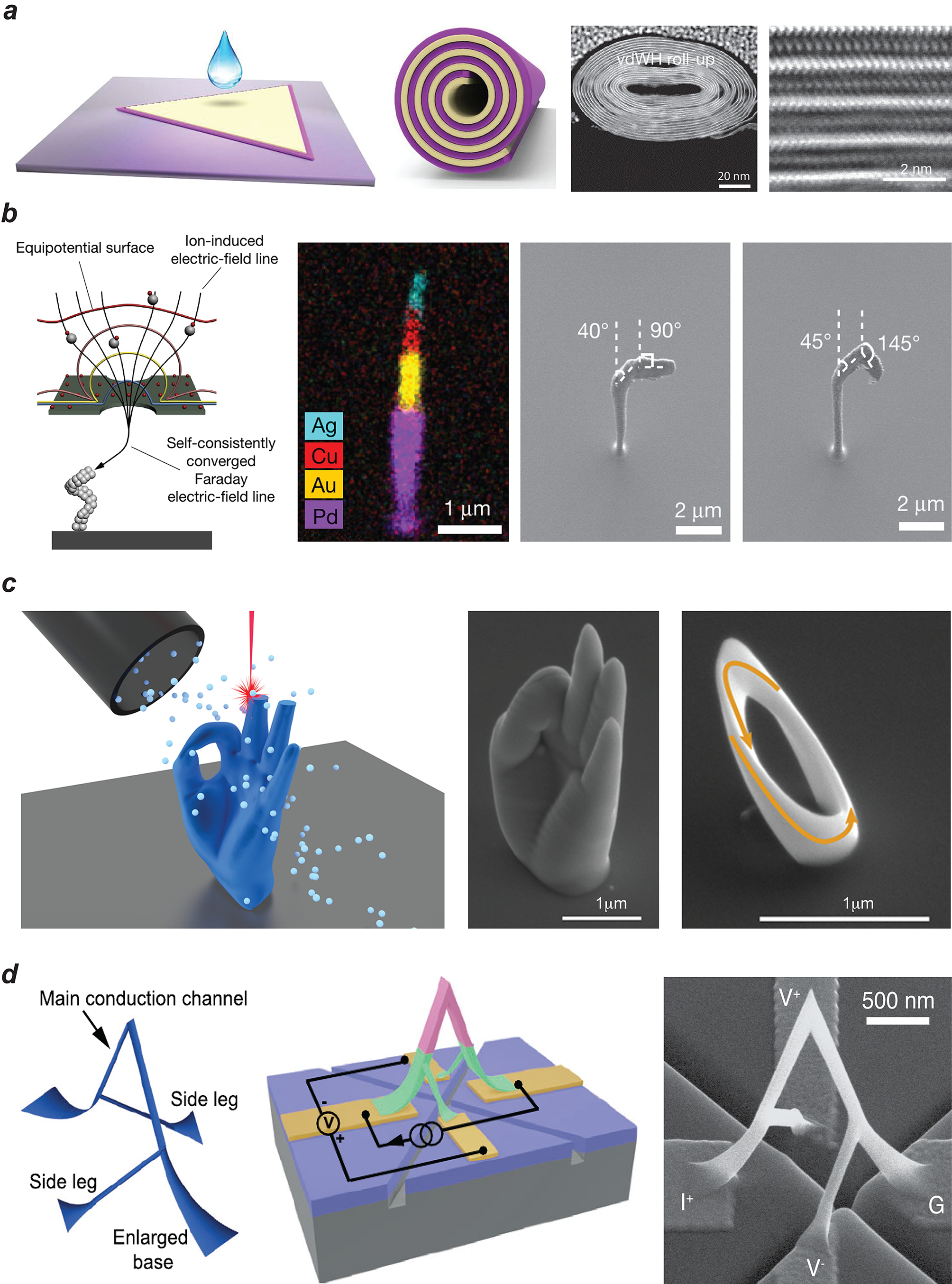}
\caption{\textbf{Fabrication of curvilinear nano-objects.} (a) Schematic illustration of the fabrication process for the formation of high-order rolled up van der Waals heterostructure. The cross-section of the SnS$_2$/WSe$_2$ superlattice was imaged by means of scanning transmission electron microscopy (TEM). Adapted  from Ref.~\cite{Zhao21a}. (b) Schematics of nanopillars fabrication by means of charged aerosol particles guided with electrostatic lens and resulting nanopillar formed with various element and geometrical composition imaged by means of scanning electron microscopy (SEM) with energy dispersive X-ray spectroscopy. Adapted  from Ref.~\cite{Jung21}. (c) Layer-by-layer focused electron beam induced deposition of various self-standing geometries with complex topographies and topologies. Adapted  from Ref.~\cite{Skoric20}.
(d) Integration of a ferromagnetic 3D nanobridge in a microelecttonic circuit by means of FEBID direct writing tecnique. Adapted  from Ref.~\cite{Meng21}.}
\label{fig4}
\end{figure*}

The use of planar structures is ideally suited to identify signatures of confinement-induced curvature effects. However, it does not give access to a number of classical shape effects and strain-driven effects that rely on the synthesis of three-dimensional nanoarchitectures. There are different routes to the fabrication of these complex nanosystems. 
The first one is based on strain engineering and involves the synthesis at some critical stage of a free-standing nanomembrane released from its substrate by a specialized anisotropic etching procedure~\cite{rog11}. 
A spatial distribution of strain, engineered in the nanomembrane for instance by heteroepitaxial growth of materials with different lattice constants, then produces a tendency to roll the layer up to create nanotubes~\cite{Schmidt01a} or coils~\cite{smi11}. More elaborate architectures can be formed using additional litographic patterning. 
The rolling-up process can be also induced externally. For instance, van der Waals planar heterostructures and monolayers can be driven into a capillary-driven roll-up~\cite{cui18,Zhao21a} by inserting or dropping ethanol solution [see Fig.\ref{fig4}(a)]. 
The formation of these 
nanoarchitectures has led to observe distinctive signatures of shape-driven curvature effects, including the linear transverse magnetoresistance mentioned above. 
Moreover, compact three-dimensional nanoarchitectures fabricated by rolled-up nanotechnology have footprints orders of magnitude smaller than conventional planar structures. 
This additional advantage has triggered interest in realizing various functional devices ranging from semiconductor electronic field effect transistors~\cite{Grimm12} and high-sensitive sensors for magnetic fields~\cite{Mueller12}, to rolled-up Josephson junctions~\cite{Thu10nal}. 

The synthesis of three-dimensional nanoarchitectures with complex shapes necessitates other strategies. For example, curved templates such as lithographically patterned substrates, spherical nanoparticles, nanocylinders, or ion-beam induced cones can be coated with a functional material. This generally yields geometric shapes with strongly inhomogeneous thicknesses relevant for shape effects in metallic materials~\cite{das19}. Patterned substrates can be used to create buckled structures with periodic strain fields that can exist over macroscopic regions~\cite{ho21}. 
Spherical geometries obtained with the use of silicon-dioxide nanoparticles has allowed to map curvature effects on the magnetic vortex states of magnetic permalloy caps~\cite{Streubel12a}. 
Another method involves the use of pre-streched elastomeric substrates in which strain relaxation imparts forces at a collection of litographically patterned locations of precursive planar structures. 
This results in a process of compressive buckling~\cite{xu15rogers} that extends the structures in the third dimension with broad geometric diversity. 

Complex three-dimensional nanosystems can be also fabricated using direct growing and writing methods. 
Glancing angle deposition, for instance, has been employed to create various functional nanoarchitectures~\cite{Zhao02,Gibbs14}. By tuning the rotation speed of the substrate, vapour source flux intensity and angle of incidence, nano-pillars, nano-flowers and nano-helix arrays have been synthesized. 
A direct coupling between the chirality of the helices and magnetism~\cite{Eslami14} has been directly proven by the occurrence of magneto-chiral dichroism of light. 
Recent direct writing techniques include two-photon lithography combined with deposition postprocessing~\cite{Williams18,Hunt20}, charged aerojet nanoprinting~\cite{Jung21}, and focused electron and ion-beam induced depositions~\cite{Teresa16,Huth18,Skoric20} [see Fig.~\ref{fig4}(b),(c)].  
Direct writing techniques allow to directly integrate complex three-dimensional architectures into micorelectronic circuitry with litographically patterned contacts [see Fig.~4(d)]. Using this method, magnetotransport studies of a ferromagnetic three-dimensional nanobridge has brought to light 
unusual angular dependences in magnetoresistive effects, including the anomalous Hall effect~\cite{Meng21}. 
For three-dimensional nanomagnets a full understanding of curvature-induced effects can be achieved by putting in a one-to-one correspondence the geometric shape with the magnetization distribution. This can be addressed using  the recently developed soft X-ray magnetic tomography~\cite{Streubel15a} and holographic vector field electron tomography~\cite{Wolf19}.

\section{Outlook} 
 These recent advances in the synthesis of three-dimensional nanosystems with complex shapes offer perspectives for the exploration of new geometry-induced nanoscale effects. For example, three-dimensional nanomagnets with a space curve geometry are expected to exhibit a torsion-induced asymmetric spin-wave dispersion~\cite{she15} that can be detected using time-resolved scanning transmission X-ray microscopy experiments. 
 When using helimagnetic materials, three-dimensional nano-objects can be also used for the design of magnetoelectric devices. Since the magnetic state can be efficiently controlled by changing the geometry of the wire, 
 embedding the nanomagnet in a piezoelectric matrix allows to achieve an electric-field induced switching between different magnetic states~\cite{vol19} that can be assigned logical `1' and `0'. 
 Three-dimensional geometries offer means to alter not only intrastructure magnetization textures~\cite{mak22} but also offer opportunities to tailor magnetic field nanotextures~\cite{don22}. Curvilinear architectures with reconfigurable magnetic field nanotextures are appealing for domain wall-based memory and logic devices: the possibility to pin magnetic domain walls in curvilinear nanowires and to tailor stray field profiles could help optimise the sensor readout of racetrack memories. Moving forward, networks of curvilinear nanowires could be employed for three-dimensional magnonics and related concepts of reservoir computing~\cite{gro20}. Curvilinear antiferromagnets will be instead central in new developments of antiferromagnetic spintronics. Recent theoretical insights promise novel means to control magnetochiral responses, induce weak ferromagnetism and tailor magnonic band gaps~\cite{pyl20}.

The principle relating strain fields and pseudomagnetic fields in Dirac materials provides a road map to exploit effective gauge fields in nanoelectronics via either spin or pseudospin control~\cite{pes12}. Challenges in the material synthesis are central in this regard. Promising directions are for instance robotic assembly of twisted van der Waals solids~\cite{man22} and controlled formation of nanobubbles via irradiation~\cite{zam15} or by trapping of substances~\cite{khe16}, in which case curvature radii down to one nm could be achieved~\cite{vil21}.
Further progress in the realization of van der Waals rolled-up tubular structures might allow to explore directional-dependent magnetotransport properties intrinsically related to the broken rotational symmetry of a spiralling roll-up. 
The anisotropy in the transversal magnetoresistance has been predicted~\cite{cha14} to decrease as $1/w$ with the number of windings of the roll $w$, and thus it is practically undetectable in multiple winding nanostructures.
Another direction to explore is the design of new material structures for the realization of the classical geometric non-linear Hall effect~\cite{sch19}. Because of its very general requirements, this phenomenon can be expected to appear full force also in traditional semiconducting materials where fabrication of high-quality planar curved channels can be achieved. 

A
frontier in the field is to individuate electronic fingerprints of the quantum geometric potential,  as its existence has been verified so far only in metamaterials~\cite{sza10}. In thin films with inhomogeneous curvature profile, the quantum geometric potential results in an in-plane internal electric field. 
This built-in electric field can potentially drive a linear Hall-like effect at zero magnetic field in materials with substantial Berry curvature such as transition metal dichalcognides. 
This Magnus Hall effect~\cite{pap19}  could therefore provide the very first experimental confirmation of the existence of the quantum geometric potential. 
The role of geometry in the family of Hall effects is expected to become even more relevant when reaching the quantum Hall regime. In (synthetic) material structures with locally non-flat geometries, the electronic density of a quantum Hall fluid is directly coupled to the Gaussian curvature via the so-called mean orbital spin~\cite{shi16}. Furthermore, additionally designing lattice disclinations would lead to an intrinsic rotation of the electronic fluid caused by gravitational anomalies~\cite{can16}. Synthesis of electronic nanomembranes with engineered nanodomes of substantial Gaussian curvature represent a promising path to bring this concept to reality. This would allow for an entirely new generation of tabletop experiments where concepts of elasticity, topology and cosmology are intertwined. Such crucial experimental efforts will also rely on new theoretical directions to understand how to disentangle physical phenomena related to confinement-induced curvature effects and geometric transport effects.

Additional evidence of the interplay  between the real-space geometry of a nanosystem and the internal geometry of the electronic wavefunctions, first unveiled in corrugated bilayer graphene~\cite{ho21},  could be pursued in complex oxides such as SrRuO$_3$. 
Ultrathin films of this magnetic material are characterized by strong Berry curvature~\cite{thi21}. Moreover, monocrystalline nanomembranes have been isolated~\cite{pas16} and could be transferred to curved templates. Periodically curved layers of this material have been also obtained starting out from a ferroelectric-metal superlattice~\cite{had21}. Oxide electronic devices thus represent an ideal material platform to explore nanoscale geometry effects in quantum phenomena. 
The use of  nanoscale curvature for the synthesis of superstructures with reduced crystalline symmetry content is likely to play a vital role in the search for unconventional superconducting phases, including orbital combinations of pair density waves~\cite{mer22}, which could be relevant for superconducting orbitronics. Crucial will be also efforts in using nanoscale geometry-induced effects to design more resilient topological superconducting phases~\cite{lae20} or even turn a ``conventional" superconductor into a topological one. We envision that developments in this direction could result in new paradigms for topological superconducting circuitry.

The recent advances, which we have reviewed, together with the new opportunities and challenges mentioned above
indicate curved electronics as a nascent and rich subject of research relevant for fundamental physics and device engineering.

 \section{Box 1: Origin of confinement-induced curvature effects} 
 Let us first illustrate the emergence of curvature effects resulting from the quantum dynamics of charge carriers in the non-relativistic regime.  
Consider an electron entering a curved channel. For its mean trajectory to follow the geometry of the structure, the electron must be subject to an usual (harmonic) confining potential with the addition of an electric field 
directed along the normal that forces its mean velocity to change direction. Clearly, the strength of this electric field grows linearly with the local curvature of the channel. To monitor its effect, it is possible to employ an adiabatic separation between fast and slow quantum degrees of freedom. In particular, the quantum motion in the strongly confined fast normal direction can be obtained by solving at each tangential position a one-dimensional Schr\"odinger equation. Due to the presence of the electric field, the energy levels departs from those of the quantum harmonic potential. This deviation is of the second-order since the first-order correction vanishes by symmetry. The local energy correction is thus negative and builds up an attractive potential for the slow degrees of freedom $\propto \hbar^2 \kappa(s) ^2 / m^{\star}$ with $\kappa(s)$ the local geometric curvature and $m^{\star}$ the electronic effective mass. 
The existence of this quantum geometric potential has been rigorously proved by Jensen, Koppe~\cite{jen71} and Da Costa~\cite{dac82} by employing a thin-wall quantization procedure. It starts with the Schr\"odinger equation in a generic curved portion of space embedding the surface or line of interest. The latter can be written as $- \hbar^2 G^{i j} {\mathcal D}_i {\mathcal D}_j \psi = 2 m^{\star} E \psi$ where one introduces the covariant derivative of a generic vector field ${\mathbf v}$  as  
$${\mathcal D}_i v_j = \partial_i v_j - \Gamma_{i j}^k  v_k,$$ 
with  $\Gamma_{i j}^k$ being the antisymmetric Christoffel symbols. The connection and the metric tensor $G_{i j}$ can be related to the geometric properties of the lower-dimensional 
manifold
assuming the quantum particle is strongly confined in the normal direction. Moreover, it is possible to safely take a zero thickness limit once a rescaled wavefunction with a well-defined surface density probability is introduced. The end product is a dimensionally-reduced Schr\"odinger equation for a free particle but with the addition of the quantum geometric potential 
$${\mathcal V}_g= -\dfrac{\hbar^2}{2 m^{\star}} \left( M^{2} - K \right),$$
where $M$ is the mean curvature and $K$ is the gaussian curvature of the generic bent surface. For the case of one-dimensional curves with a single principal curvature $\kappa(s)$ the geometric potential reduces to $-\hbar^2 \kappa(s)^2 / ( 8 m^{\star})$ in agreement with the qualitative argument given above. 

Materials with sizable spin-orbit coupling possess additional geometry-induced effects. Consider for instance a curvilinear one-dimensional nanostructure. Its low-energy effective Rashba spin-orbit coupled Hamiltonian can be written as 
$${\mathcal H}= -\dfrac{\hbar^2}{2 m^{\star}} \partial_s^2 + \dfrac{i \hbar \alpha}{2} \left[ \sigma_N(s) \partial_s + \partial_s \sigma_N(s) \right],$$ 
 where $\alpha$ is the Rashba spin-orbit coupling strength and we introduced the triad of Pauli matrices $\left\{ \sigma_T, \sigma_N, \sigma_B \right\}$ comoving with the curvilinear Frenet-Serret frame  determined by the tangential, normal, and binormal directions~\cite{ort15}. 
 By completing the square in the Hamiltonian, it is possible to show that the expectation values of the spin component are determined by a local operator $\hat{G}(s)= - \sigma_N(s) / (2 l_{\alpha}) - \beta~\sigma_0$
where $l_{\alpha}=\hbar / (2 m^{\star} \alpha)$ is the spin-orbit interaction length while $\beta$ is a constant that depends on the eigen-energy and does not affect the spin textures. By further using the commutation relations for the triad of Frenet-Serret of Pauli matrices, 
one finds
an equation that links the electron spin orientation in the Frenet-Serret frame, the strength of the Rashba spin-orbit interaction and the geometric curvature. This fundamental equation~\cite{yin16} reads 
$$\partial_s \braket{\boldsymbol \sigma} = - {\bf h}_{eff} \times \braket{\boldsymbol \sigma}, $$ 
where the effective field lies in the normal-binormal plane and has an explicit geometric component given by the local curvature $\kappa(s)$ since ${\bf h}_{eff} = \left\{0, l_{\alpha}^{-1} , \kappa(s) \right\}$. 
For non-zero curvature the electron spin acquires a finite out-of-plane binormal component. Additionally, and as discussed in the main text, for inhomogeneous curvature there is a finite torque that yields an unconventional tangential spin component. 
This curvature control of the spin textures is reflected in the real-space geometry control of a quantum geometric phase. There is in fact a direct relation linking the spin texture to the Aharonov-Anandan (AA) geometric phase. We recall that the AA phase is the non-adiabatic analog of the Berry phase which can be defined as $\gamma=\displaystyle \oint_ {\mathcal C} d {\bf r} \cdot {\mathcal {A}}$, with ${\mathcal A}$ the Berry connection
${\mathcal A}= -i  \bra{ \psi({\bf r})} \nabla \ket{\psi({\bf r})}$. The Berry curvature is the field strength with components $\Omega_{r_a}=\epsilon_{a b c} \partial_{r_b} {\mathcal A}_{r_c}$ and transforms as a pseudovector in three-dimensions, and as a pseudoscalar in two-dimensional systems.

In ferromagnetic materials, confinement effects on the exchange magnetic energy yield a curvature-induced Dzyaloshinskii-Moriya interaction (DMI) and a magnetic anisotropy. 
Precisely as for the quantum geometric potential, one starts with the exchange energy in a thin curved shell with the energy density written as usual ${\mathcal E}_{ex}= {\boldsymbol \nabla} {\bf m} \cdot {\boldsymbol \nabla} {\bf m}$,  where ${\bf m}$ is the 
magnetization. The zero-thickness limit of this three-dimensional exchange magnetic energy yields a surface energy consisting of three-different terms~\cite{kra16}, {\it i.e.} ${\mathcal E}_{ex} = {\mathcal E}_{ex}^0  + {\mathcal E}_{ex}^D + {\mathcal E}_{ex}^A$.  For a generic curved surface and assuming an orthonormal curvilinear local basis has been found, the three different contributions can be written as follows 
\begin{eqnarray*} 
{\mathcal E}_{ex}^0&=& \nabla_{\alpha} m_{\beta} \nabla_{\alpha} m_{\beta} + \nabla_{\alpha} m_{n} \nabla_{\alpha} m_{n}  \\
{\mathcal E}_{ex}^D&=& 2 h_{\alpha \beta} \left( m_{\beta} \nabla_{\alpha} m_{n} - m_{n} \nabla_{\alpha} m_{\beta} \right) + 2 \epsilon_{\alpha \beta} \Omega_{\gamma} m_{\beta} \nabla_{\gamma} m_{\alpha}  \\
{\mathcal E}_{ex}^A&=& \left(h_{\alpha \gamma} h_{\gamma \beta} + \Omega^2 \delta_{\alpha \beta} \right) m_{\alpha} m_{\beta} + \left( M^2 - 2 K \right) m_n^2 + \\ & & 2 \epsilon_{\alpha \gamma} h_{\gamma \beta} \Omega_{\beta} m_{\alpha} m_n . 
\end{eqnarray*} 
In the equation above, we introduced the Weingarten curvature tensor $h_{\alpha \beta}$, the spin connection $\Omega$ associated of the two-dimensional curved surface, and the two-dimensional Levi-Civita tensor $\epsilon_{\alpha \beta}$. Additionally, $m_n$ is the component of the 
magnetization in the direction normal to the curved surface. 
The emergence of the curvature-induced DMI interaction, in particular, reflects the fact that bending of a curved thin magnetic layer breaks the centrosymmetry. 

In superconductors, 
an homogeneous spin-triplet pairing order parameter can be conveniently expressed 
introducing the so-called $\bm{d}$-vector:  
\begin{equation*}
\widehat{\Delta}=
\left(\begin{array}{cc}\Delta^{\uparrow\uparrow} & \Delta^{\uparrow\downarrow}\\\Delta^{\uparrow\downarrow} & \Delta^{\downarrow\downarrow}\end{array}\right) \equiv
i \left(\bm{d}\cdot\bm{\sigma} \right) \sigma_y, 
\label{DeltaT}
\end{equation*}
where we used the relation $\Delta^{\uparrow\downarrow}=\Delta^{\downarrow\uparrow}$. 
The complex components of ${\bm d}$-vector are related to the pair amplitudes 
by
\begin{equation*}
\bm{d}=\left(
-\frac{\Delta^{\uparrow\uparrow}-\Delta^{\downarrow\downarrow}}{2},\frac{\Delta^{\uparrow\uparrow}+\Delta^{\downarrow\downarrow}}{2i},\Delta^{\uparrow\downarrow}\right)\,.
\end{equation*}
 Hence, each component of the $\bm{d}$-vector indicates the pair amplitude for the Cooper-pair spin perpendicular to the corresponding spin axis. 
Non-vanishing amplitudes of the $\bm{d}$-vector components are mainly dictated by the structure of the pairing potential while the orientation of the $\bm{d}$-vector is determined, for instance, by spin-orbit coupling, magnetic fields or intrinsic magnetism. 
Triplet pairing with a non-zero value of the product $\bm{d}\times \bm{d}^{*}$ implies that the spins of the Cooper pairs are polarized. 
Additionally, 
the relative $0$ ($\pi$) spin-phase difference appearing between the $\Delta^{\uparrow\uparrow}$ and $\Delta^{\downarrow\downarrow}$ matrix elements when only the $d_y$ ($d_x$) component is present, can be relevant for superconducting spintronic applications.  
One can generally show that there will be a non-trivial Josephson coupling, both in the charge and spin channel, which depends on the relative $\bm{d}$-vector misalignment angle. For the case of a typical Josephson junction configuration with tunnel coupled superconducting regions marked by $\bm{d}$-vectors with a misalignment angle $\beta$, the Josephson current~\cite{asa06} for the charge and spin sector is proportional to $\left[\sin(\phi+\beta)+\sin(\phi-\beta)\right]$  and $\left[\sin(\phi+\beta)-\sin(\phi-\beta)\right]$ respectively, with $\phi$ being the phase difference between the $\bm{d}$-vectors.

 \section{Box 2: Gauge fields in strained nanostructures} 
We illustrate the emergence of gauge fields in strained nanosystem with Dirac electrons by considering the specific example of graphene, and start with the simplest tight-binding model Hamiltonian, {\it i.e.} considering only hopping processes between nearest neighbor atomic sites, which reads  
\begin{equation*}
{\mathcal H}_{MLG}= -  \sum_{i \, n} t_n a^{\dagger}_i b_{i + \delta_n} + {\it c.c.}, 
\end{equation*} 
where $a^{\dagger}$ and $b^{\dagger}$ ($a$, $b$) are creation (annihilation) operators on the A and B sublattices respectively. In the equation above, the subscript $i$ runs over all unit cell positions, and we introduced the three nearest neighbor vectors
\begin{equation*}
{\boldsymbol \delta}_{1}=\dfrac{a}{\sqrt{3}} \left\{\dfrac{\sqrt{3}}{2}, \dfrac{1}{2} \right\} ; {\boldsymbol \delta}_2= \dfrac{a}{\sqrt{3}}  \left\{\dfrac{-\sqrt{3}}{2}, \dfrac{1}{2} \right\} ; {\boldsymbol \delta}_3 = \dfrac{a}{\sqrt{3}} \left\{0, -1 \right\}. 
\end{equation*} 

The presence of strain implies that the hopping amplitudes $t_n$ explicitly depend on the nearest neighbor vectors. Specifically,
$t_n = t_0 \left( 1 - \beta \, \delta u_n \right)$
where $t_0$ is the hopping amplitude for the pristine threefold rotation symmetric honeycomb lattice, the lattice parameter $\beta$ can be determined by Raman spectroscopy whereas the relative distance changes $\delta u_n$ can be expressed in terms of the strain tensor components $\epsilon_{i\, j}$ as 
\begin{equation*}
\delta u_n = \dfrac{\delta_n^i \, \delta_n^j}{a^2} \epsilon_{i j}.
\end{equation*}
 To proceed further, we go to momentum space and write the Bloch Hamiltonian as
\begin{equation*} 
{\mathcal H}_{MLG} = - \sum_n t_n \left( \begin{array}{cc} 0 & e^{-i \left({\bf K}^{(\prime)} + {\bf q} \right) \cdot {\boldsymbol \delta}_n} \\  e^{i \left({\bf K}^{(\prime)} + {\bf q} \right) \cdot {\boldsymbol \delta}_n}  & 0
 \end{array} \right), 
\end{equation*}
where we have rewritten the momenta as ${\bf k} = {\bf K}^{(\prime)} + {\bf q}$ since we are interested in the electronic properties close to the ${\bf K}$ or ${\bf K}^{\prime}$ valleys of the Brillouin zone (BZ) given by ${\bf K}= \left\{ \dfrac{4 \pi}{3 \sqrt{3} a} , 0 \right\}$ and ${\bf K}^{\prime}= \left\{-\dfrac{4 \pi}{3 \sqrt{3} a} , 0 \right\}$. 
The Bloch Hamiltonian can be then expanded to linear order in the small momenta $q$. Using simple vector identities and assuming for simplicity an anisotropic biaxial strain with $\epsilon_{xx} \neq \epsilon_{yy} \neq 0$ and $\epsilon_{x y} \equiv 0$ the continuum low-energy Hamiltonian near the two valleys of the BZ can be then recast as 
\begin{equation*}
\mathcal{H}_{eff}(\mathbf{q})= \xi v_x q_x\sigma_x+v_F {\mathcal A}_x\sigma_x + v_y q_y \sigma_y 
\end{equation*}
where $\xi=\pm 1$ is the valley index. 
Here ${\mathcal A}_x = \sqrt{3}\, \beta(\epsilon_{xx}-\epsilon_{yy})/ (2 a)$ is a  strain-induced ``pseudo"-gauge field  whereas $v_F=\sqrt{3}t_0 a / 2$  is the Fermi velocity of the Dirac carriers in unstrained samples. In addition, $v_{x}= v_F[1-\beta  (3\epsilon_{xx}+\epsilon_{yy})/4]$ and $v_y=v_F[1-\beta (\epsilon_{xx}+3\epsilon_{yy})/4]$ are renormalized Fermi velocities that become anisotropic due to momentum-strain coupling.

{\bf Acknowledgements:}
We would like to acknowledge the numerous colleagues with whom we have collaborated on the topics described in this Review. 
We would also like to acknowledge the Future and Emerging Technologies (FET) Programme within the Seventh Framework Programme for Research of the European Commission under FET-Open Grant No.618083 (CNTQC) for the financial support to the work performed by the authors on this topic. 
C.O. acknowledges support from a VIDI grant (Project 680-47-543) financed by the Netherlands Organization for Scientific Research (NWO). The work of D.M. and O.M.V. was financed in part via numerous national and European projects including German Research Foundation (DFG) Grants MA 5144/9-1, MA 5144/13-1, MA 5144/28-1, VO 2598/1-1, and Helmholtz Association of German Research Centres in the frame of the Helmholtz Innovation
Lab "FlexiSens". Z.-J. Y. acknowledges support from the National Natural Science Foundation of China (Grant No. 11974151.)

{\bf Author Contributions:}
C.O. coordinated the project. P.G. produced the original illustrations. All authors wrote and commented on the manuscript.

\end{document}